\documentstyle[prl,aps,floats,preprint,psfig]{revtex}   % preprint
\begin{document}
\draft

%%%%%%%%%%%%%%%%%%%%%%%%%%%%%%%%%%%%%%%%%%%%%%%%%%%%%%%%%%%%%%%%%%%%%%%%%%
%% Macros
%%%%%%%%%%%%%%%%%%%%%%%%%%%%%%%%%%%%%%%%%%%%%%%%%%%%%%%%%%%%%%%%%%%%%%%%%%

%%%%%%%%%%%%%%%%%%%%%%%%%%%%%%%%%%%%%%%%%%%%%%%%%%%%%%%%%%%%%%%%%%%%%%%%%%
%% Title page
%%%%%%%%%%%%%%%%%%%%%%%%%%%%%%%%%%%%%%%%%%%%%%%%%%%%%%%%%%%%%%%%%%%%%%%%%%

%\Draft                 % For "Draft" diagonal annotation in galley mode
%\Preprint              % For "Preprint" diagonal annotation in galley mode

\title{Chain Diffusion in Lamellar Block Copolymers}

\author{Michael Murat$^1$\cite{murataddress},
Gary S.\ Grest$^2$, and Kurt Kremer$^1$}

\address{
$^1$Max Planck Institut f\"ur Polymerforschung,
Postfach 3148, 55021 Mainz, Germany}

\address{$^2$Corporate Research Science Laboratories,
Exxon Research and Engineering Company,
Annandale, New Jersey 08801}

\date{\today}

\maketitle
{
\begin{abstract}
Diffusion of symmetric diblock copolymer chains in macroscopically oriented
lamellar block copolymers are studied in a molecular dynamics
simulation. Results for diffusion constant both parallel $D_\parallel$ and
perpendicular  $D_\perp$ to the lamellar planes
are compared to results in the disordered one phase region. For diblocks of
length $N=40$ ($1.1N_e)$ and $100\ (3N_e)$,
where $N_e$ is the entanglement length of a
homopolymer melt at the same density, $D_\parallel$ is
nearly independent of $\chi$,
while $D_\perp$ is strongly suppressed
as $\chi$ is increased.  These results are
in agreement with  theoretical predictions based
on the Rouse model.
%For quenched disordered
%systems, the isotropic $D\simeq (2D_\parallel+D_\perp)/3$ of the
%corresponding lamellar system.
The isotropic diffusion constant of quenched disordered systems is
approximated fairly well by $(2D_\parallel+D_\perp)/3$ of the corresponding
lamellar system.
\end{abstract}
}
{
\pacs{\mbox{ }}
}

\narrowtext

%%%%%%%%%%%%%%%%%%%%%%%%%%%%%%%%%%%%%%%%%%%%%%%%%%%%%%%%%%%%%%%%%%%%%%%%%%
%% Core Text
%%%%%%%%%%%%%%%%%%%%%%%%%%%%%%%%%%%%%%%%%%%%%%%%%%%%%%%%%%%%%%%%%%%%%%%%%%

The question whether lamellar ordering modifies self diffusion in a symmetric
diblock melt has been the subject of several recent theoretical and
experimental studies. Barrat and Fredrickson\cite{barrat91} argued that for
chains obeying Rouse dynamics, the diffusion within a lamellar plane is
unaffected by the lamellar structure, while perpendicular diffusion is
retarded by an amount determined by the degree of segregation. Measurements on
quenched, unoriented samples \cite{shull91a,dalvi93b}
indicated that there is no discontinuity for the diffusion constant
at the order-disorder transition (ODT).
The  samples used in these studies were
polycrystalline, without macroscopic ordering, so that the separate
components of the diffusion coefficient could not be determined. Such a
separation was achieved\cite{dalvi93a} for  entangled chains by
macroscopically orienting the sample using oscillatory shear. Measurements  on
such mono
domain samples showed less anisotropy than predicted by the Rouse model. For
well entangled chains, both parallel and perpendicular
diffusion  have to involve translation of the A segments of the chains into the
B domains, so that both components are reduced by the lamellar structure,
though not necessarily to the same extent. Such an activated reptation
mechanism in which the diffusion constant $D$ decays as $\exp(-\chi N)$ was
observed by Lodge and Dalvi\cite{lodge95} for highly entangled melts of
poly(ethylenepropylene)-poly(ethylethylene) diblock copolymers. The
difference between the diffusion constants parallel $D_\parallel$
and perpendicular $D_\perp$ to the interface was found to decrease as the ODT
is approached. Block-retraction mechanism, similar to arm-retraction known
from star polymers, is also found to play a role in the diblock diffusion and
also gives rise to an exponential decay of $D$ with increasing
$N$.

Computer simulations of diblock copolymer systems have been mostly restricted
to studying the statics and thermodynamics of such systems
\cite{binder95}.
The only simulation which studied diffusion
in a diblock system is  by Pan and Shaffer\cite{pan96},
who studied a symmetric diblock in the strong segration regime
($\chi N=45$)  for several chain lengths $N$
using the bond fluctuation method on a lattice.
They found that
$D_\parallel$ was
roughly two times smaller than $D$ for a homopolymer melt and the
critical chain length at which entanglements became important was
comparable in the two systems.
No perpendicular
diffusion was observed for this value of $\chi N$.
Since these simulations were done on a lattice, there was
no {\it a priori} way to determine the equilibrium lamellar spacing $d_l$.

Here we study by continuum space molecular dynamics simulations
the diffusion constants of perfectly oriented
lamellar systems both in the plane of the lamella and in the perpendicular
direction, for chains of length 40 and 100,
corresponding to about 1.1 and $3N_e$ respectively.
$N_e\simeq 35$ is the entanglement length for a homopolymer
melt \cite{kremer90}, at the same average density ($\rho\simeq
0.85\sigma^{-3}$). $d_l$ is determined from a constant pressure
simulation.  It is then fixed and $D_\parallel$ and $D_\perp$ are
determined at constant volume. The results are compared to simulations in the
isotropic disordered phase.

We use a coarse grained bead-spring model similar to that
used in our earlier studies of polymer melts and networks
\cite{kremer90}
and tethered chains \cite{grest95}.
Each chain  consists of $N$ beads (monomers), connected to form a linear
chain.  For diblocks, the model is generalized to
two types of polymer species, say $A$ and $B$,
which are connected in   a block of $fN$ beads of
type $A$ connected to a block of $(1-f)N$ beads of type $B$.
Here we  consider only symmetric diblocks, $f=1/2$.
The interaction potential $U_{IJ}(r)$ between two beads
of types $I, J = \{A, B\}$ separated by a distance
$r$ is taken as the repulsive part
of a Lennard Jones $6:12$ potential,
\begin{equation}
U_{IJ}(r)=
\cases{
        4\epsilon_{IJ} \left[ \left(\frac{\sigma_{IJ}}{r}\right)^{12}
                -\left(\frac{\sigma_{IJ}}{r}\right)^6 + \frac{1}{4} \right]
        & $r \le r_c$;\cr
        0
        & $r > r_c$, \cr
}
\label{LJ}
\end{equation}
where the cutoff distance $r_c = 2^\frac{1}{6}\sigma_{IJ}$. Here,
$\epsilon_{IJ}$ and $\sigma_{IJ}$ are,
respectively, parameters fixing the
energy and length scale for monomers of  type $I$ and $J$.
Adjacent monomers along the chains are coupled through an anharmonic
FENE potential, chosen to eliminate unphysical
bond crossings or breaking \cite{kremer90,grest95}.
All the interactions are short ranged and the model is very
efficient computationally. For symmetric copolymers, we set
$\epsilon_{AA} = \epsilon_{BB} = \epsilon,$
$\epsilon_{AB}/\epsilon = 1 + \tilde{\epsilon},$
and $\sigma_{AA} = \sigma_{BB} = \sigma_{AB} = \sigma.$
The latter choice simply means that all monomers have the
same interaction range and thus the two species have the same
molar volume.
All of our results are reported in terms of
$\sigma$ and  $\epsilon$, with $\tau = \sigma (m/\epsilon)^{1/2}$
for the time scale, where $m$ is the mass of a monomer.
The energy parameters
are a special case of ($\epsilon = 1$)
${\tilde\epsilon}=\epsilon_{AB}-\frac{1}{2}(\epsilon_{AA}+\epsilon_{BB})$,
used in simple lattice models such as in Flory-Huggins theory.
In ref. \cite{grest96c}, we showed that simply
introducing  this difference in the
repulsive interaction strength between like and unlike monomers
was sufficient to drive phase separation.
By varying ${\tilde\epsilon}$, while leaving the temperature $T$ constant,
one can induce a phase transition
from a homogeneous blend to a system with
two segregated coexisting phases $[{\tilde\epsilon}_c=3.40(5)k_BT/N]$ 
\cite{grest96c}.
This approach also has the advantage of overcoming a problem inherent in
experimental studies: namely, the dependence of the monomeric friction
constant on the temperature. It is somewhat reminiscent of
recent elegant experiments by Gehlsen {\it et al.}\cite{gehlsen92} in which
the phase separation of binary mixtures
is studied as a function of the difference
in deuterium content between two otherwise identical polymer species.

\begin{figure}[tb]
\centerline{
\psfig{figure=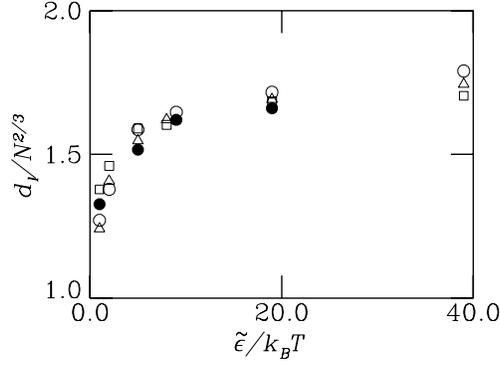,height=2.0in,%
bbllx=70pt,bblly=190pt,bburx=440pt,bbury=470pt%
}}
\caption{The equilibrium lamellar spacing $d_l$
scaled by $N^{2/3}$ as a function of
the excess AB interaction parameter ${\tilde\epsilon}$, for 
$N=40\ (\bigtriangleup)$,
$100\ (\circ)$, $200\ (\Box)$, and $400\ (\bullet)$.
\label{f:dl}}
\end{figure}

We study $M$ chains of length $N$ by solving Newton's equations of motion
using a velocity-Verlet algorithm
\cite{allen87} with a time step $\Delta t=0.013\tau$
\cite{grest95,grest96c}.  Two
different starting states were used: disordered and lamellar. In the former
case, we studied cubic systems at density $\rho=
0.85\sigma^{3}$\cite{kremer90}. The
lamellar state was constructed as described in ref.~\cite{grest96c}. Since
$d_l$ depends on  ${\tilde\epsilon}$ and is not
known {\it a priori}, we first ran a constant pressure simulation
\cite{grest96c} at a $P=5\epsilon\sigma^{-3}$ in which the dimensions
of the simulation cell, parallel $L_\parallel$ and perpendicular $L_\perp$
to the lamellar plane varied independently \cite{parrinello80}
for each value of ${\tilde\epsilon}$. $d_l$ adjusts itself
rather rapidly (a few hundred thousand time steps). With
increasing ${\tilde\epsilon}$,   $L_\perp$ increases
and the cross sectional area $L_\parallel^2$
decreases, such that the overall
density remains fairly constant.
The equilibrium lamellar spacing
increases rapidly with ${\tilde\epsilon}$ and then saturates as shown in Fig.\
\ref{f:dl}.   The results scale with $N$
as predicted by self-consistent field  theory
\cite{helfand80,semenov85,ohta86}.
For large ${\tilde\epsilon}$
the chains are highly stretched and the number of
$AB$ contacts is small. Thus, further increase in the repulsive energy
between the two species does not strongly affect $d_l$.
In this regime $\chi$ is no longer proportional to ${\tilde\epsilon}$
as often incorrectly assumed in many previous lattice
simulations of diblock copolymer systems \cite{binder95,pan96}.

After $d_l$ is determined, we fixed the volume to
study the dynamics in the lamellar phase. The results presented here
are for $M=1600$ chains of length $N=10$,
$M=800$  chains of $N=20$ and $40$ and $M=200$ chains of
$N=100$. The $N=40$ and $100$ systems were studied in both the
disordered and lamellar regions of phase space, while
$N=10$ and $20$ systems were only simulated
in the disordered phase. The  lamellar
systems contained 100 chains per layer for $N=40$
and $50$ chains per layer for $N=100$, giving $L_\perp=4d_l$ for $N=40$
and $2d_l$ for $N=100$.  From our study of the order parameter in
the ordered phase as a function of ${\tilde\epsilon}$, we estimate that
${\tilde\epsilon}_{ODT}(40)=
0.85\pm 0.05$ and ${\tilde\epsilon}_{ODT}(100)=0.28\pm 0.03$ \cite{grest96c}.
To study the diffusion in both phases, long runs were necessary
to obtain good statistics. Typical runs for $N=40$ were of
length $(5.2-10.0)\times 10^4\tau$  while for $N=100$,
they were of order $(3.9-6.5)\times 10^5\tau$. In addition, lamellar systems
for $N=200$ and $N=400$ were also
simulated  for visualization of the chain  motion.
However, these simulations were too large to  run long enough
to obtain the  diffusive constants.  These results will be presented in a
later paper.

We follow the motion of the chains' center of mass through $g_3(t)$, whose
components are defined by
\begin{equation}
g_{3\alpha}(t)=\left\langle \left[r_{{\rm cm}\alpha}(t)-r_{{\rm cm}\alpha}(0)
\right]^2 \right\rangle ,
\label{g3alpha}
\end{equation}
with $\alpha=x$, $y$, or $z$. Here $r_{{\rm cm}\alpha}(t)$ are the components
of the
center of mass of a chain and  $\langle\cdots\rangle$ denotes both an
ensemble
average over all the chains in the system and an average over different
starting states. At sufficiently long times, $g_{3\alpha}(t)$
exhibits diffusive motion with $g_{3\alpha}\rightarrow 2D_\alpha t$.
In the disordered phase,
the three diffusion constants are identical and equal to $D$, the
isotropic diffuion constant.
In the lamellar phase, the motion in the
$xy$ plane (plane of the lamella) is expected to be faster than the motion
between the lamellar planes ($z$ direction). The parallel
diffusion constant $D_\parallel=(D_x+D_y)/2$, and
$D_\perp=D_z$.

\begin{figure}[tb]
\centerline{
\psfig{figure=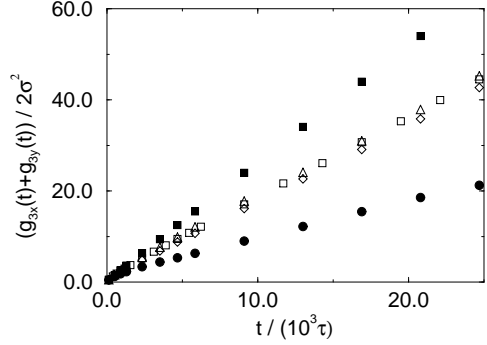,height=2.2in}
}
\caption{
Mean squared diplacement of the center of mass parallel to the lamellar plane,
$[(g_{3x}(t)+g_{3y}(t))/2]$ versus $t/\tau$ for $N=40$ systems: lamellar with
${\tilde\epsilon}=0.9\ (\bigtriangleup$) and $3.0\ (\diamond)$, homopolymer melt
(${{\vrule height .9ex width .8ex depth -.1ex }}$),
isotropic near the ODT, ${\tilde\epsilon}=0.8 (\Box)$, and at 
${\tilde\epsilon}=3.0 (\bullet)$ .
\label{f:g3n40}}
\end{figure}

\begin{figure}[tb]
\centerline{\psfig{figure=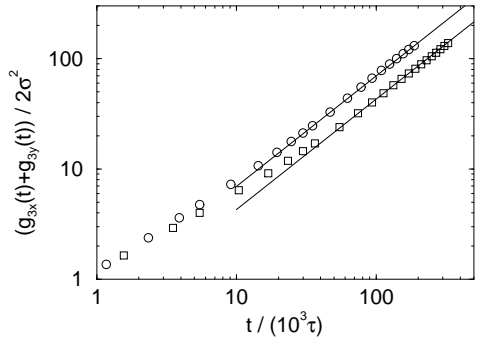,height=2.2in}
}
\caption{$(g_{3x}(t)+g_{3y}(t))/2$ versus  $t/\tau$ for a homopolymer melt
($\circ$) and a lamellar system with ${\tilde\epsilon}=1.0\ (\Box$) for
$N=100$.
The lines have a slope of 1, characteristic of the diffusive regime.
\label{f:g3n100}}
\end{figure}

The behavior of $(g_{3x}(t)+g_{3y}(t))/2$ for some of the $N=40$ systems
is shown in Fig.~\ref{f:g3n40}. For the isotropic systems
(homopolymers and the disordered diblocks), the quantity shown is $g_3(t)/3$.
For times longer than about $10^4\tau$, all the systems show diffusive
behavior. The same quantity for a homopolymer melt and a lamellar system of
chain length $N=100$ is shown
in Fig.~\ref{f:g3n100}.
The transition to the diffusive regime in the lamellar system takes place
around $t \approx 10^4\tau$, somewhat later than in the homopolymer
melt. This delay in the onset of the diffusive motion is common to
all lamellar systems studied. The diffusion constants were
calculated from the slope of $(g_{3x}(t)+g_{3y}(t))/2$ in the diffusive
regime. Perpendicular diffusion is observed only for systems sufficiently
close to the ODT. For large ${\tilde\epsilon}$, $g_{3z}(t)$
saturates at about the square of the thickness of the interface region, within
which the junction points of the chains are localized.

\begin{figure}[tb]
\centerline{\psfig{figure=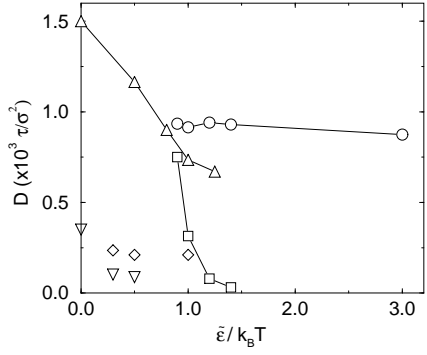,height=2.0in}
}
\caption{
Diffusion coefficients as a function of the excess AB interaction parameter,
${\tilde\epsilon}$:
$D_\parallel \ (\circ)$ and $D_\perp \ (\Box)$ for $N=40$ lamella, isotropic
$D \ (\bigtriangleup)$ for disordered  $N=40$ melt, $D_\parallel \ (\diamond)$
for $N=100$ lamella and isotropic $D \ (\bigtriangledown)$ for disordered
$N=100$ melt.  The lines for the $N=40$ systems are only a guide to the eye.
\label{f:difn40}}
\end{figure}

Chains of length 40 $(1.1N_e)$ are clearly in the Rouse regime and are
expected to obey Rouse dynamics. Barrat and Fredrickson modelled the self
diffusion of a Rouse chain in a lamellar system as tracer difusion of such a
chain in a periodic
potential varying in one dimension, perpendicular to the lamella
planes\cite{barrat91}.
According to this study, diffusion in the plane is not affected by the
periodic potential, while diffusion perpendicular to the lamella is slowed
down by an amount dependent upon the magnitude of the concentration
fluctuations. Near the ODT, $D_\perp$
is expected to decrease by
less than 40\% compared to $D_\parallel$,
while in the strongly segregated phases it should decay
exponentially with  $\chi N$\cite{barrat91}. To check these predictions, we
plot in Fig.~\ref{f:difn40} both $D_\parallel$ and $D_\perp$ as a function 
of ${\tilde\epsilon}$ for lamellar systems as well as the isotropic $D$
for disordered systems. $D_\parallel$ in
the lamellar $N=40$ systems is only very weakly dependent on ${\tilde\epsilon}$. 
At the
ODT, $D_\parallel\tau/\sigma^2 \simeq 9.5 \times 10^{-4}$, while in the strong
segregration limit (${\tilde\epsilon}=9.0$,
not shown in Fig.~\ref{f:difn40}), $D_\parallel\tau/\sigma^2$ decreases
only  to $8.0 \times 10^{-4}$.
For a homopolymer melt $({\tilde\epsilon}=0)$, the isotropic
diffusion  constant is about $1.5\times10^{-3}\sigma^2/\tau$, so that the
diffusion constant in the isotropic phase close to ODT is lower by
about 50\% compared to the diffusion constant in the homopolymer melt. This
reduction results from the increased friction from the local motion of the
monomers. $D_\perp$, on the other hand, decays rapidly with ${\tilde\epsilon}$.
While near the ODT the ratio  $D_\perp/D_\parallel\simeq 0.8$, it is
reduced to about 0.03 for ${\tilde\epsilon}=1.4$. 
The reduction of about 20\% near ODT is
consistent with the Barrat-Fredrickson prediction. For higher values of
${\tilde\epsilon}$, we could not observe any perpendicular diffusion 
during the time scale of our simulations.

The four non-zero values of $D_\perp$ for $N=40$ can be fitted nicely with an
exponential
decay of the form $D_\perp /D_o = 0.57\exp
[-0.16N({\tilde\epsilon}-{\tilde\epsilon}_{ODT})]$,
where $D_o$ is the  value the diffusion constant in the homopolymer melt.
To make a quantitative comparison with
the theory of  Barrat and Fredrickson \cite{barrat91}, we note that they
predict that when $\chi
\psi_0\approx \chi_{ODT}$, where $\psi_0$ is the amplitude of the density
variation in the lamella, $D_\perp$ is reduced by a factor of about 10. For
our system at ${\tilde\epsilon}=1.0>{\tilde\epsilon}_{ODT}$, $\psi_0=0.83$, 
so that ${\tilde\epsilon}\psi_0=0.83\approx {\tilde\epsilon}_{ODT}$. 
For this system, $D_\perp$ is about $1/3$
of the ODT value or about 3 times less suppression than predicted.

The situation for the $N=100$ chains is qualitatively similar. This
length corresponds to about $3N_e$, and entanglements are not expected to
play a major role. In a homopolymer of $N=100$ chains, crossover to reptation
motion is barely observed \cite{kremer90}. For stretched chains, $N_e$ might
even increase.
The systems near the ODT all have essentially the
same $D_\parallel$ (about $2.1\times 10^{-4}\sigma^2/\tau$).
The reduction in $D$ at the ODT compared to the homopolymer melt is
about $40\%$, comparable to that seen for $N=40$.
For very high values of ${\tilde\epsilon}$, the reduction in $D_\parallel$ is
comparable to that for $N=40$ chains.
In the most
strongly stretched system studied (${\tilde\epsilon}=9.0$, not shown in
Fig.~\ref{f:difn40}), $D_\parallel$ is about 0.75 of
$D_\parallel$ near ODT, while the same ratio is about 0.85 for $N=40$.
The perpendicular motion is
significantly reduced in comparison to the parallel one.
For ${\tilde\epsilon}=0.3$ (close to the ODT), we found $D_\perp=7\times
10^{-5}\sigma^2/\tau$, or about $1/3$ of the corresponding
$D_\parallel$.  We note that the suppression of $D_\perp$ near ODT is higher
than for the $N=40$ diblocks and the Barrat-Fredrickson prediction. However,
for these longer chains, a
more precise determination of the ODT is needed before a reliable
comparison can be made.
For larger values of ${\tilde\epsilon}$, no measurable perpendicular
diffusion was found.  For the system with ${\tilde\epsilon}=0.5$, only
two chains succeeded in translating their junction points from one
interface to another during the whole course of the simulation, with no
measurable $D_\perp$. For larger ${\tilde\epsilon}$,
no chain exchange occurred between the interfaces.

It is also interesting to note that for quenched disordered systems (those
with ${\tilde\epsilon} > {\tilde\epsilon}_{ODT}$), the isotropic $D\simeq
(2D_\parallel+D_\perp)/3$ of the corresponding lamellar system\cite{comment1}.
Fredrickson\cite{fredrickson93a} had
suggested that for small anisotropy, the diffusion
constant  of a quenched
system should be slightly larger than $(2D_\parallel+D_\perp)/3$. We do not
see an evidence of the predicted enhancement. However, this prediction is only
for systems very close to the ODT. Since the exact
ODT of our model system is not known to a high accuracy, we cannot rule out
this result.

We also studied the diffusion in the disordered regime for shorter,
unentangled chains
of length $N=10$ and $20$.  For $N=10$ and $20$,
the isotropic diffusion constant scales $D\sim N^{-1} 
f({\tilde\epsilon}/{\tilde\epsilon}_{ODT})$,
where $f(x)$ is a decreasing funciton of $x$.
A  comparision of these results
with the theoretical predicitions  of Leibig and Fredrickson \cite{leibig96}
for tracer diffusion in the disordered regime will be presented in a later
paper.

Our results demonstrate that for lamellar systems of short chains (of length
$N=40$ and $100$),
self diffusion in the lamellar plane near the ODT
is essentially unaffected by the
amount of chain stretching. This is in accordance with the Rouse-like motion
of chains. Perpendicular diffusion is strongly suppressed by the lamellar
structure. Our results are qualitatively consistent with the Barrat-Fredrickon
theory, although the amount of suppression is somewhat less than predicted. In
the limit of strong stretching, some reduction
in $D_\parallel$ is observed, possibly indicating the onset of
entanglement effects. Further studies on long chains are needed to
study the effect of entanglements.

We thank T.~P.\ Lodge for helpful discussions. This work was
supported in part by the BMBF project ``Computer Simulations of
Complex Materials".

%%%%%%%%%%%%%%%%%%%%%%%%%%%%%%%%%%%%%%%%%%%%%%%%%%%%%%%%%%%%%%%%%%%%%%%%%%
%% References
%%%%%%%%%%%%%%%%%%%%%%%%%%%%%%%%%%%%%%%%%%%%%%%%%%%%%%%%%%%%%%%%%%%%%%%%%%
%\bibliographystyle{../prsty}
%\bibliography{../macros,../ref13,notes}

\begin{thebibliography}{10}
\bibitem{comment1}
Disordered systems simulated at ${\tilde\epsilon}>{\tilde\epsilon}_{ODT}$ 
are not in equilibrium, and
should eventually evolve into lamellar systems over time scales much longer
than the simulation time. The measured $D$ of such systems are derived from
the effective slope of $g_3(t)$. Experimentally quenched systems are
polycrystalline, and the diffusion in such systems is affected to a large
extent by the morphology.
\end{thebibliography}

%%%%%%%%%%%%%%%%%%%%%%%%%%%%%%%%%%%%%%%%%%%%%%%%%%%%%%%%%%%%%%%%%%%%%%%%%%
%% Figures
%%%%%%%%%%%%%%%%%%%%%%%%%%%%%%%%%%%%%%%%%%%%%%%%%%%%%%%%%%%%%%%%%%%%%%%%%%

\end{document}